\documentclass{PoS}

\usepackage{amsmath}

\title{Neutral pion form factor measurement at NA62}

\ShortTitle{Neutral pion form factor measurement at NA62}

\author{\speaker{Evgueni Goudzovski}\thanks{Supported by ERC Starting Grant 336581. For the NA48/2 and NA62 collaborations:
F.~Ambrosino, A.~Antonelli, G.~Anzivino, R.~Arcidiacono,
W.~Baldini, S.~Balev, J.R.~Batley, M.~Behler, S.~Bifani, C.~Biino
% ...
}\\
        School of Physics and Astronomy\\University of Birmingham, B15 2TT, United Kingdom\\
        E-mail: \email{eg@hep.ph.bham.ac.uk}}

\abstract{The NA62 experiment at CERN collected a large sample of charged kaon decays with a highly efficient trigger for decays into electrons in 2007. The kaon beam represents a source of tagged neutral pion decays in vacuum. A measurement of the electromagnetic transition form factor slope of the neutral pion in the time-like region from $1.05\times10^6$ fully reconstructed $\pi^0$ Dalitz decay is presented. The limits on dark photon production in $\pi^0$ decays from the earlier kaon experiment at CERN, NA48/2, are also reported.}

\FullConference{38th International Conference on High Energy Physics\\
		3-10 August 2016\\
		Chicago, USA}

\begin{document}

\section*{Introduction}

The NA48/2 experiment at the CERN SPS collected a large sample of charged kaon decays in 2003--04 (corresponding to about $2\times 10^{11}$ $K^\pm$ decays in the vacuum decay volume). The experiment used simultaneous $K^+$ and $K^-$ beams and was optimized for the search for direct CP violating charge asymmetries in the $K^\pm\to3\pi^\pm$ decays~\cite{ba07}. Its successor, the NA62-$R_K$ experiment, collected a 10 times smaller $K^\pm$ decay sample with low intensity beams and minimum bias trigger conditions in 2007 using the same detector~\cite{la11,la13}. The large data samples accumulated by the two experiments have allowed precision studies of rare $K^\pm$ decays and searches for new physics. The $K^\pm$ beams represent a copious source of tagged neutral pions produced and decaying in vacuum. This paper is focused on neutral pion physics at NA48/2 and NA62-$R_K$: a measurement of the slope of the $\pi^0$ electromagnetic transition form factor from a sample of $\pi^0\to e^+e^-\gamma$ decays and a search for the dark photon ($A'$) in the $\pi^0\to\gamma A'$ decay are reported.

\section{Beam and detector}
\label{sec:experiment}

The NA48/2 and NA62-$R_K$ experiments used simultaneous $K^+$ and $K^-$ beams produced by 400~GeV/$c$ primary CERN SPS protons impinging on a beryllium target. Charged particles in a narrow momentum band were selected by an achromatic system of four dipole magnets which split the two beams in the vertical plane and recombined them on a common axis. The beams then passed through collimators and a series of quadrupole magnets, and entered a 114~m long cylindrical vacuum tank with a diameter of 1.92 to 2.4~m containing the fiducial decay region.

The vacuum tank was followed by a magnetic spectrometer housed in a vessel filled with helium at nearly atmospheric pressure, separated from the vacuum by a thin ($0.3\%~X_0$) $\rm{Kevlar}\textsuperscript{\textregistered}$ window. An aluminium beam pipe of 158~mm outer diameter traversing the centre of the spectrometer (and all the following detectors) allowed the undecayed beam particles to continue their path in vacuum. The spectrometer consisted of four drift chambers (DCH) with an octagonal transverse width of 2.9~m: DCH1, DCH2 located upstream and DCH3, DCH4 downstream of a dipole magnet that provided a horizontal transverse momentum kick of 120~MeV/$c$ for charged particles (for NA48/2). Each DCH was composed of eight planes of sense wires. The DCH space point resolution was 90~$\mu$m in both horizontal and vertical directions, and the momentum resolution was $\sigma_p/p = (1.02 \oplus 0.044\cdot p)\%$, with $p$ expressed in GeV/$c$ (for NA48/2). The spectrometer was followed by a plastic scintillator hodoscope (HOD) with a transverse size of about 2.4 m, consisting of a plane of vertical and a plane of horizontal strip-shaped counters arranged in four quadrants (each logically divided into four regions). The HOD provided time measurements of charged particles with 150~ps resolution. It was followed by a liquid krypton electromagnetic calorimeter (LKr), an almost homogeneous ionization chamber with an active volume of 7 m$^3$ of liquid krypton, $27~X_0$ deep, segmented transversally into 13248 projective $\sim\!2\!\times\!2$~cm$^2$ cells. The LKr energy resolution was $\sigma_E/E=(3.2/\sqrt{E}\oplus9/E\oplus0.42)\%$, the spatial resolution for an isolated electromagnetic shower was $(4.2/\sqrt{E}\oplus0.6)$~mm in both horizontal and vertical directions, and the time resolution was $2.5~{\rm ns}/\sqrt{E}$, with $E$ expressed in GeV. The calorimeter was followed by a muon system consisting of three scintillator planes, with an iron wall installed in front of each plane. A detailed description of the beamline, detector and trigger is given in Refs.~\cite{ba07,la13,fa07}.

\boldmath
\section{$\pi^0$ transition form factor slope measurement}
\unboldmath
\label{sec:pi0dalitz}
As a $\pi^0$ is produced in four of the six main $K^\pm$ decays, the NA62-$R_K$ experiment exposed to about $2\times10^{10}$ kaon decays in flight in the vacuum fiducial decay region with minimum bias trigger conditions is an ideal environment to study the neutral pion physics. The Dalitz decay $\pi^0_D\to\gamma e^+ e^-$ proceeds through the $\pi^0\gamma\gamma$ vertex with an off-shell photon. The commonly used kinematic variables defined in terms of the
particle four-momenta are:
\begin{equation*}
    x = \left( \frac{M_{e e}}{m_{\pi^0}} \right)^2
    = \frac{(p_{e^+} + p_{e^-})^2}{ m_{\pi^0}^2}, \qquad
    y = \frac{2 \, p_{\pi^0} \cdot \left( p_{e^+} - p_{e^-} \right)}{m_{\pi^0}^2
    (1-x)},
\end{equation*}
where $p_{\pi^0}$, $p_{e^+}$, $p_{e^-}$ are respectively the $\pi^0$ and $e^\pm$ four-momenta, $m_{\pi^0}$ is the mass of the $\pi^0$, and $M_{ee}$ is the $e^+e^-$ invariant mass. The physical region is given by
\begin{equation*}
    r^2 = \left(\frac{2 m_e}{m_{\pi^0}}\right)^2 \leq x \leq 1, \quad |y| \leq
    \sqrt{1 - \frac{r^2}{x}} \;.
\end{equation*}
The $\pi^0_D$ differential decay width normalised to the $\pi^0_{2\gamma}\to\gamma\gamma$ decay width is
\begin{equation*}
    \frac{1}{\Gamma(\pi^0_{2\gamma})} \frac{\text{d}^2 \Gamma(\pi^0_D)}{\text{d}x \text{d}y} =
    \frac{\alpha}{4\pi} \frac{(1-x)^3}{x} \left(1 + y^2 + \frac{r^2}{x}\right) \; \left(1+\delta(x,y)\right) \;
    \left|\mathcal{F}(x)\right|^2,
\end{equation*}
where $\mathcal{F}(x)$ is the semi-off-shell electromagnetic transition form factor (TFF) of the $\pi^0$ and the $\delta(x,y)$ term encodes the radiative corrections.

The TFF is usually expanded as $\mathcal{F}(x) = 1+ax$, where $a$ is the form factor slope parameter. This approximation is justified by the smallness of this parameter. In the vector meson dominance (VMD) model, it is dominated by the $\rho$ and $\omega$ mesons, resulting in a slope of $a\approx m^2_{\pi^0} \left(m_\rho^{-2}+m_\omega^{-2}\right)/2 \approx 0.03$. 

A crucial aspect of measuring the $\pi^0$ TFF with the $\pi^0_D$ decays is the modelling of the radiative corrections in the differential rate: their effect on the differential decay rate exceeds that of the TFF slope. The first study of the corrections on the differential rate was performed in Ref.~\cite{Lautrup1971} in the soft-photon approximation, and extended in Ref.~\cite{Mikaelian1972}. A recent improved computation~\cite{Husek2015_rad} triggered by and used for this analysis includes additional second-order contributions and accounts for the emission of radiative photons, i.e. the internal bremsstrahlung contribution.

\begin{figure}[h]
\begin{center}
\resizebox{0.47\textwidth}{!}{\includegraphics{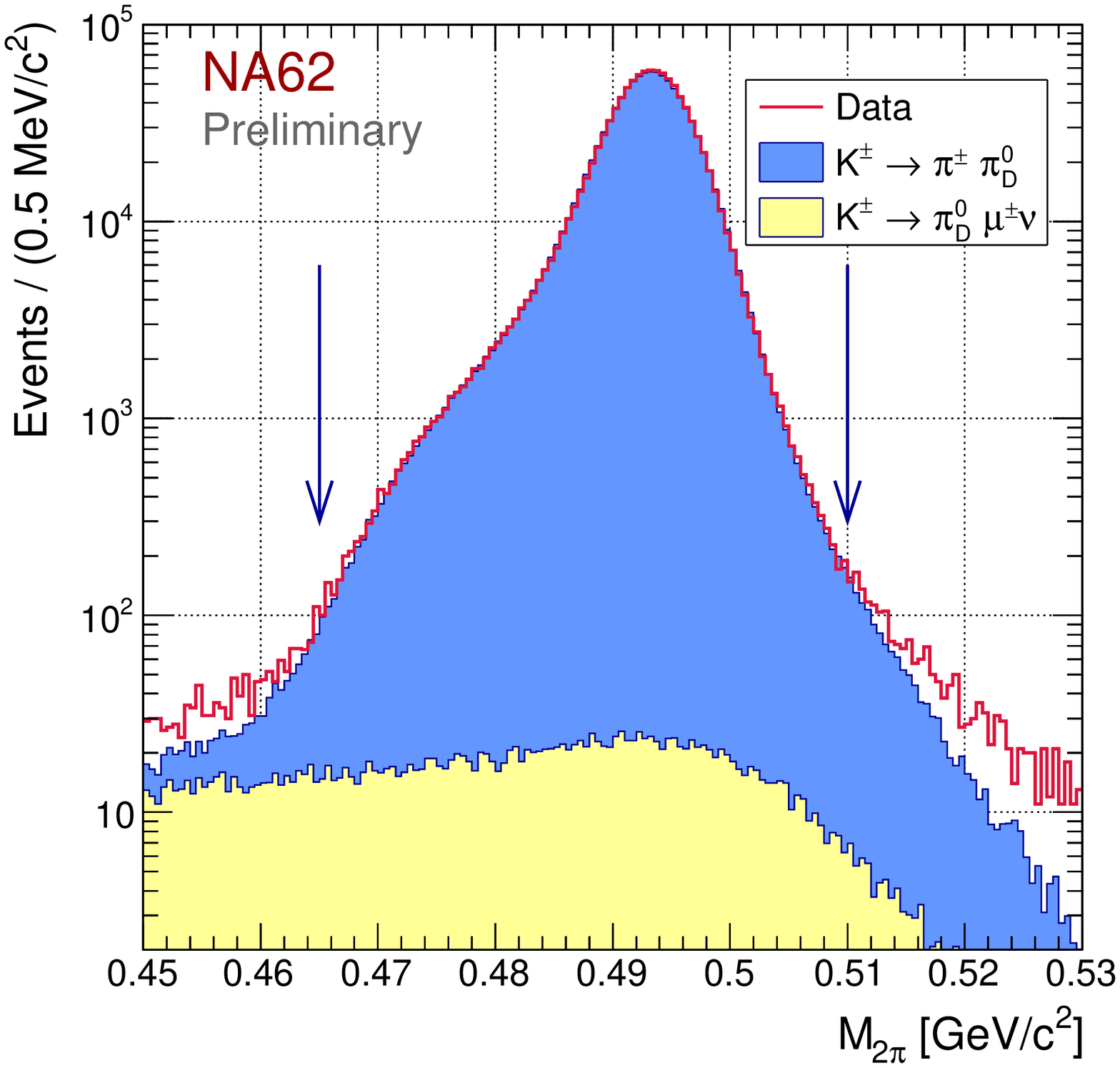}}%
\resizebox{0.53\textwidth}{!}{\includegraphics{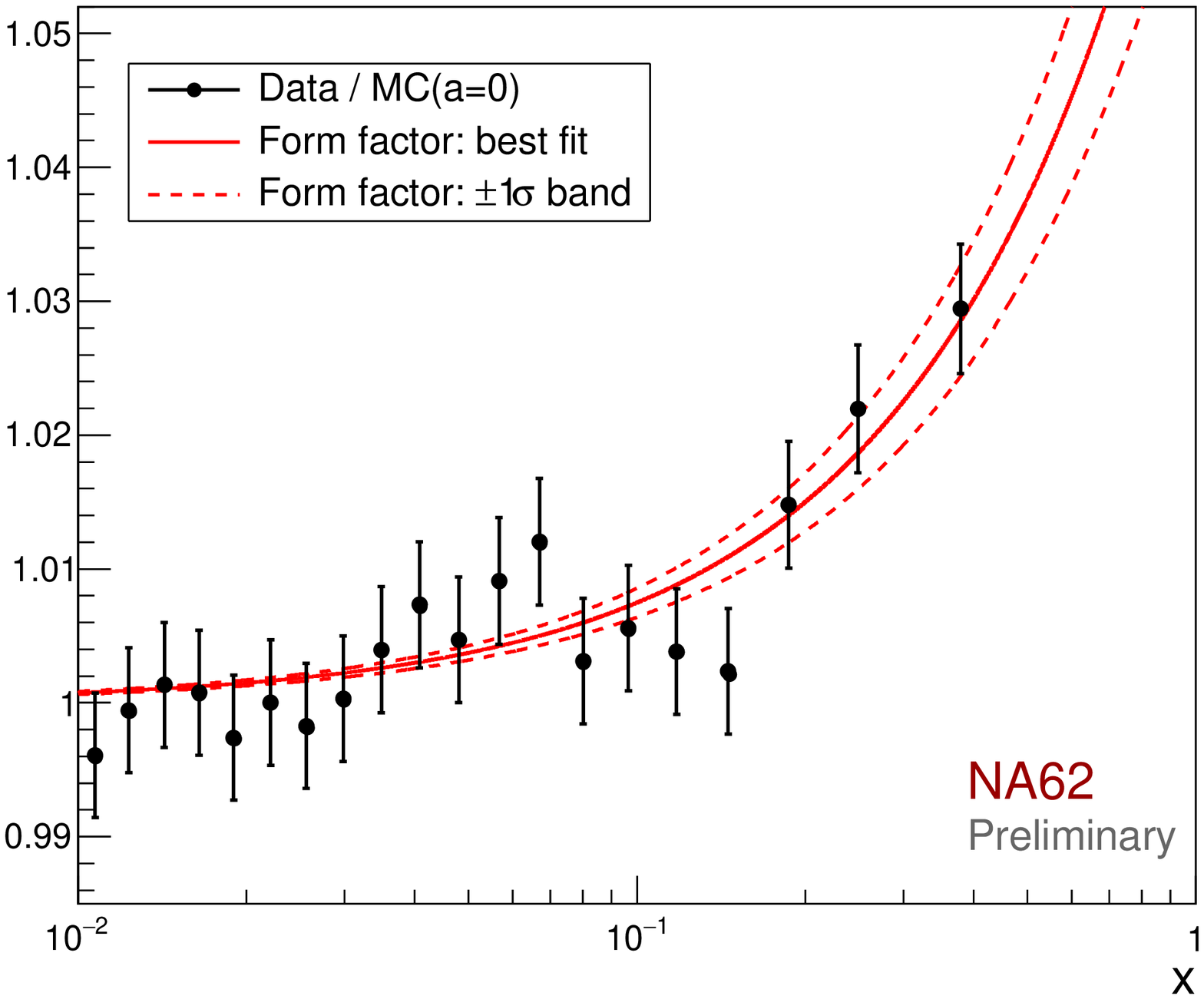}}
\end{center}
\vspace{-5mm}
\caption{Left: reconstructed $\pi^\pm\pi^0_D$ mass spectrum of the data and simulated samples. The residual difference is mainly due to the non-inclusion of the $K^\pm\to\pi^0_D e^\pm\nu$ component of the signal into the simulation. Right: Illustration of $\pi^0$ the TFF fit result showing the data/MC ratio, with the MC sample weighted to obtain $a=0$. The events are divided into 20 equipopulous bins, and the markers are located at the bin barycentres of each bin. The solid line represents the TFF function with a slope value equal to the fit central value. The dashed lines indicate the $1\sigma$ band.}
\label{fig:pi0d_result}
\end{figure}

A sample of $1.05\times 10^6$ $K^\pm\to\pi^\pm\pi^0_D$ decays with negligible background has been selected in the NA62-$R_K$ data using a spectrometer three-track vertex selection and reconstructing the photon in the LKr calorimeter. A $\chi^2$ fit to the reconstructed $x$ distribution of the data $\pi^0_D$ candidates with equipopulous binning has been performed to extract the TFF slope value (Fig.~\ref{fig:pi0d_result}). The main systematic uncertainties arise from the simulation of the beam spectrum and from the calibration of the spectrometer global momentum scale. The preliminary result is
\begin{equation*}
a = (3.70 \pm 0.53_\text{stat} \pm 0.36_\text{syst})\times 10^{-2},
\end{equation*}
with $\chi^2/\text{ndf} = 52.5/49$ corresponding to a \textit{p}-value of 0.34. This measurement represents an observation of a positive $\pi^0$ electromagnetic TFF slope (with more than $5\sigma$ significance) in the time-like region of momentum transfer.

\boldmath
\section{Search for the dark photon in the $\pi^0\to\gamma A'$ decay}
\unboldmath

A search for a hypothetical dark photon (DP, denoted $A'$) using a large sample of tagged $\pi^0$ mesons from reconstructed $K^\pm\to\pi^\pm\pi^0$ and $K^\pm\to\pi^0\mu^\pm\nu$ decays is reported below; the details are available for Ref.~\cite{ba15}.

In a rather general set of hidden sector models with an extra $U(1)$ gauge symmetry~\cite{ho86}, the interaction of the DP with the visible sector proceeds through kinetic mixing with the Standard Model (SM) hypercharge. Such scenarios with GeV-scale dark matter provide possible explanations to the observed rise in the cosmic-ray positron fraction with energy and the muon gyromagnetic ratio $(g-2)$ measurement~\cite{po09}. The DP is characterized by two a priori unknown parameters, the mass $m_{A'}$ and the mixing parameter $\varepsilon^2$. Its possible production in the $\pi^0$ decay and its subsequent decay proceed via the chain $\pi^0\to\gamma A'$, $A'\to e^+e^-$. The expected branching fraction of the above $\pi^0$ decay is~\cite{batell09}
\begin{equation*}
{\cal B}(\pi^0\to\gamma A') = 2\varepsilon^2 \left(1-\frac{m_{A'}^2}{m_{\pi^0}^2}\right)^3 {\cal B}(\pi^0\to\gamma\gamma),
\label{eq:br}
\end{equation*}
which is kinematically suppressed as $m_{A'}$ approaches $m_{\pi^0}$. In the DP mass range $2m_e<m_{A'}<m_{\pi^0}$ accessible in pion decays, the only allowed tree-level decay into SM fermions is $A'\to e^+e^-$. Therefore, for a DP decaying only into SM particles, ${\cal B}(A'\to e^+e^-)\approx 1$, and the expected total decay width is~\cite{batell09}
\begin{equation*}
\Gamma_{A'} \approx \Gamma(A'\to e^+e^-) = \frac{1}{3} \alpha\varepsilon^2 m_{A'} \sqrt{1-\frac{4m_e^2}{m_{A'}^2}}\left(1+\frac{2m_e^2}{m_{A'}^2}\right).
\end{equation*}
It follows that, for $2m_e\ll m_{A'}<m_{\pi^0}$, the DP mean proper lifetime $\tau_{A'}$ satisfies the relation
\begin{equation*}
c\tau_{A'} = \hbar c / \Gamma_{A'} \approx 0.8~{\mu\rm m} \times \left(\frac{10^{-6}}{\varepsilon^2}\right) \times \left(\frac{100~{\rm MeV}/c^2}{m_{A'}}\right).
\end{equation*}
The analysis is performed assuming that the DP decays at the production point (prompt decay), which is valid for sufficiently large values of $m_{A'}$ and $\varepsilon^2$. In this case, the DP production and decay signature is identical to that of the Dalitz decay $\pi^0_D\to e^+e^-\gamma$, which therefore represents an irreducible but well controlled background and determines the sensitivity.

\begin{figure}[tb]
\begin{center}
\resizebox{0.5\textwidth}{!}{\includegraphics{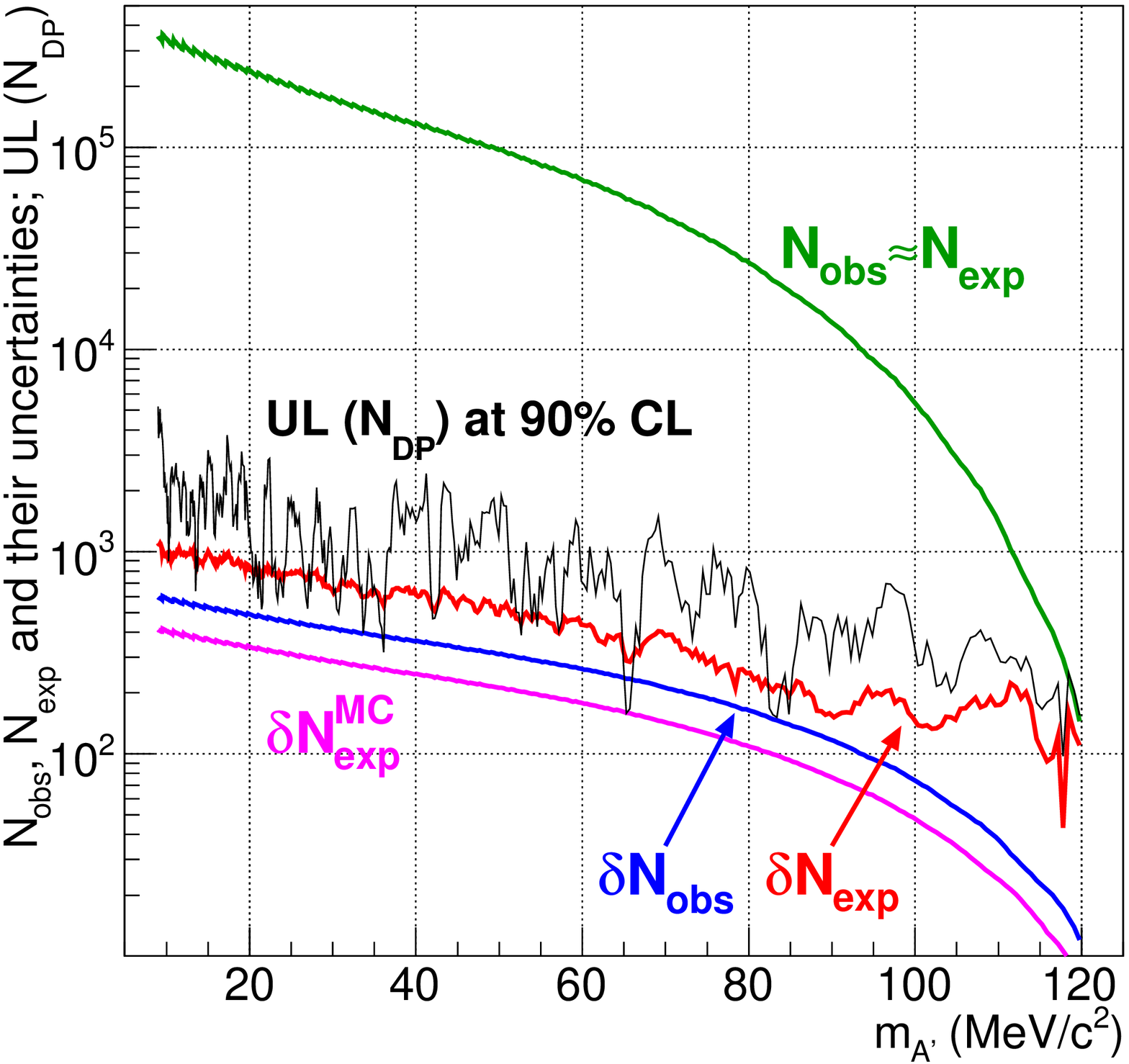}}%
\resizebox{0.5\textwidth}{!}{\includegraphics{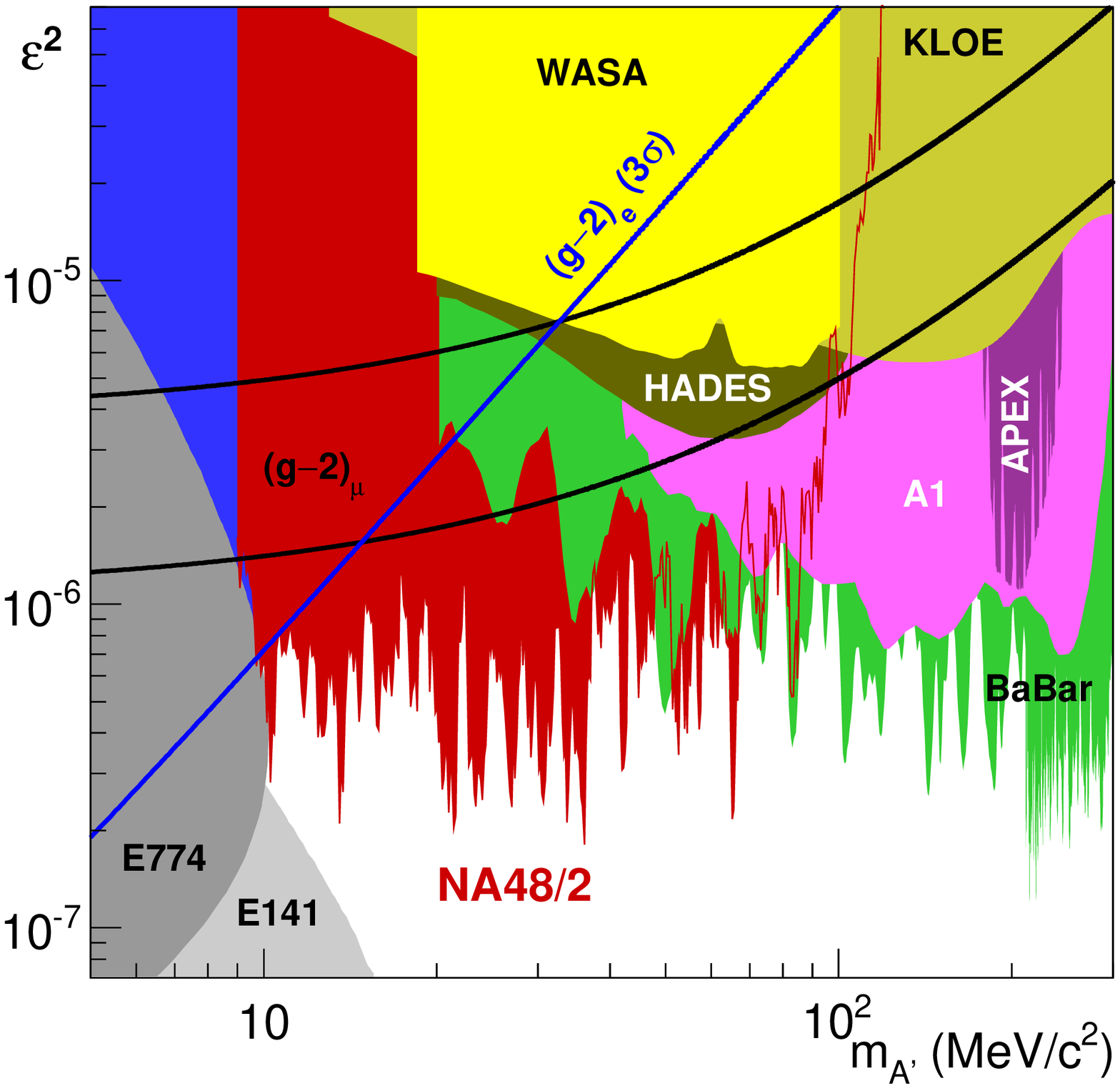}}%
\end{center}
\caption{Left: Numbers of observed data events ($N_{\rm obs}$) and expected $\pi^0_D$ background events ($N_{\rm exp}$) passing the joint DP selection (indistinguishable in a logarithmic scale), estimated uncertainties $\delta N_{\rm obs}=\sqrt{N_{\rm exp}}$ and $\delta N_{\rm exp}$, and obtained upper limits at 90\% CL on the numbers of DP candidates ($N_{\rm DP}$) for each DP mass value $m_{A'}$. The contribution to $\delta N_{\rm exp}$ from the MC statistical uncertainty is shown separately ($\delta N_{\exp}^{\rm MC}$). The remaining and dominant component is due to the statistical errors on the trigger efficiencies measured in the DP signal region. Right: Obtained upper limits at 90\% CL on the mixing parameter $\varepsilon^2$ versus the DP mass $m_{A'}$, compared to other published exclusion limits from meson decay, beam dump and $e^+e^-$ collider experiments. Also shown is the band where the inconsistency of theoretical and experimental values of muon $(g-2)$ reduces to less than 2 standard deviations, as well as the region excluded by the electron $(g-2)$ measurement. Further details can be found in Ref.~\cite{ba15}.}
\label{fig:observed}
\end{figure}

The full NA48/2 data set is analyzed, and two sources of $\pi^0_D$ decays are considered: $K^\pm\to\pi^\pm\pi^0$ and $K^\pm\to\pi^0 e^\pm\nu$ decays followed by the prompt $\pi^0\to\gamma A'$, $A'\to e^+e^-$ decay chain. The two event selections are identical up to the momentum, invariant mass and particle identification conditions. The total number of reconstructed $\pi^0_D$ candidates is $1.69\times 10^7$, with a negligible background contamination. A scan for a DP signal in the mass range $9~{\rm MeV}/c^2 \le m_{A'} < 120~{\rm MeV}/c^2$ has been performed. The lower boundary of the mass range is determined by the limited accuracy of the $\pi^0_D$ background simulation at low $e^+e^-$ mass. At high DP mass approaching the upper limit of the mass range, the sensitivity to the mixing parameter $\varepsilon^2$ is not competitive with the existing limits due to the kinematic suppression of the $\pi^0\to\gamma A'$ decay. The numbers of observed events and expected background events in the signal $e^+e^-$ mass window for each DP mass hypothesis considered, as well as the obtained upper limits on the numbers of DP candidates in each mass hypothesis considered are presented in Fig.~\ref{fig:observed}. It can be seen that the sensitivity is limited by the irreducible background from $\pi^0_D$ decays.

The obtained upper limits at 90\% CL on ${\cal B}(\pi^0\to\gamma A')$ are of the order of $10^{-6}$ for the whole range of $A'$ masses considered. The computed limits on the the mixing parameter $\varepsilon^2$ for each DP mass value are shown in Fig.~\ref{fig:observed}, together with the constraints from the SLAC E141 and FNAL E774, KLOE, WASA, HADES, A1, APEX and BaBar experiments. Also shown is the band in the ($m_{A'}$, $\varepsilon^2$) plane where the discrepancy between the measured and calculated muon $(g-2)$ values falls into the $\pm2\sigma$ range due to the DP contribution, as well as the region excluded by the electron $(g-2)$ measurement. The obtained limits are more stringent than the previous ones in the mass range 9--70 MeV/$c^2$, reaching $2\times 10^{-7}$. In combination with other experimental searches, this result rules out the DP as an explanation for the muon $(g-2)$ measurement under the assumption that the DP couples to quarks and decays predominantly to SM fermions.

\section*{Conclusions}

Kaon decay in flight experiments are exposed to large numbers of tagged neutral pion decays, which makes them an appropriate environment for $\pi^0$ decay studies. A new precise measurement of the $\pi^0$ transition form factor slope has been performed by analyzing the NA62-$R_K$ data set: a positive form factor slope is observed. Improved limits on dark photon production in the $\pi^0\to\gamma A'$ decay have been obtained from the analysis of the NA48/2 data.

\end{document}